\documentclass[12pt]{article}

\usepackage{graphicx}
\usepackage[crop=pdfcrop]{pstool}
\usepackage{float}
\usepackage{siunitx}
\usepackage{chet}

\date{December 2012}

\preprint{UCSD-PTH-12-08}

\title{Limit Cycles in Four Dimensions}

\author{Jean-Fran\c{c}ois Fortin, Benjam\'\i{}n Grinstein and Andreas
Stergiou\emails{jffortin@physics.ucsd.edu, bgrinstein@ucsd.edu,
stergiou@physics.ucsd.edu}}

\affiliation{Department of Physics, University of California, San Diego, La
Jolla, CA 92093 USA}

\abstract{We present an example of a limit cycle, i.e., a recurrent
flow-line of the beta-function vector field, in a unitary four-dimensional
gauge theory. We thus prove that beta functions of four-dimensional gauge
theories do not produce gradient flows. The limit cycle is established in
perturbation theory with a three-loop calculation which we describe in
detail.}

\DeclareMathOperator{\Tra}{Tr}
\DeclareMathOperator{\Rea}{Re}
\DeclareMathOperator{\Ima}{Im}

\begin{document}

\maketitle

\newsec{Introduction}
A necessary prerequisite for the complete understanding of quantum field
theory (QFT) is the appreciation of its possible phases.  In some cases a
phase may be out of direct computational reach, e.g., the confining phase
of QCD, while in others one may be able to use perturbation theory to gain
an understanding of the dynamics of the theory. For a long time the only
perturbatively accessible phase of QFT has been presumed to be that of a
theory at a conformal fixed point, where, e.g., correlators exhibit
power-law scaling.

Recently, the existence of renormalization-group (RG) limit cycles was
established by us in $d=4-\epsilon$ spacetime dimensions with a three-loop
calculation in a unitary theory of scalars and fermions
\cite{Fortin:2011ks, Fortin:2011sz, Fortin:2012ic}. Theories in
$d=4-\epsilon$ are of course unphysical, but working with them has always
been useful in the study of properties of the RG \cite{Wilson:1973jj}, in
the sense that RG effects found in such theories have invariably been shown
to have counterparts in more physical cases.  It was therefore suggested by
our results that limit cycles should also occur in integer spacetime
dimensions.  In the present note we show that this is indeed the case in a
four-dimensional unitary gauge theory.

This new feature of the RG gives rise to an obvious question: ``what phase
of QFT is described by a limit cycle?'' It follows from the work of Jack
and Osborn \cite{Jack:1990eb} that theories that live on limit cycles may
be CFTs.  As we show in \cite{Fortin:2012hn} this is indeed the case for
the limit cycle we present below. Thus, although beta functions admit limit
cycles, theories that live on these cycles are conformal.

The existence of recurrent trajectories in the RG has implications for the
$c$-theorem. This theorem reflects the intuition that coarse-graining
reduces the number of massless degrees of freedom of a QFT, and it comes in
different versions, as explained, e.g., in \cite{Barnes:2004jj}. The strong
version, i.e., that there exists a scalar function of the couplings $c$,
along any RG flow, that obeys $dc/dt\le0$, with $t$ the RG time and the
inequality saturated only at fixed points, was proved long ago for QFTs in
$d=2$ \cite{Zamolodchikov:1986gt}, and has been elaborated on heavily in
the literature.  Soon thereafter it was suggested that a strong $c$-theorem
should be true in $d=4$ as well \cite{Cardy:1988cwa}, and that was indeed
shown to be the case at weak coupling \cite{Osborn:1989td, Jack:1990eb}, at
least when renormalization effects of certain composite operators are not
of relevance \cite{Osborn:1991gm, Fortin:2012ic}. A proof of the
four-dimensional version of the weak version of the $c$-theorem was
recently claimed \cite{Komargodski:2011vj} (see also
\cite{Komargodski:2011xv}), i.e., that there is a $c$-function such that if
two four-dimensional CFTs are connected by an RG flow, then
$c_{\text{UV}}>c_{\text{IR}}$.  Similar ideas were used in an attempt for a
proof of the weak version of the $c$-theorem in $d=6$ \cite{Elvang:2012st}.

We hasten to remark that the existence of limit cycles in the beta-function
vector field does not contradict intuition derived from the $c$-theorem.
In particular, the quantity $c$ that satisfies a $c$-theorem is constant
even on limit cycles, and is expected to have the same monotonic behavior
when it flows from a UV fixed point or limit cycle to an IR fixed point or
limit cycle. However, the existence of RG limit cycles obviously
demonstrates that beta-function flows are not gradient flows.

The outline of the paper is as follows. In the next section we present our
example. We describe in detail the three-loop calculation that establishes
the limit cycle, and we show that the dilatation current of the theory on
the limit cycle is well-defined and free of anomalies. In the last section
we conclude and mention a few open questions.

\newsec{The 4d example}[example]
In this section we describe in detail the first example of a limit cycle in
$d=4$.

\subsec{The theory}
Our theory has an $SU(3)$ gauge group with two singlet real scalars,
$\phi_1$ and $\phi_2$, two pairs of fundamental and antifundamental active
Weyl fermions, $(\psi_{1,2},\tilde{\psi}_{1,2})$, as well as
$\tfrac12(29-3\varepsilon)$ pairs of fundamental and antifundamental
sterile Weyl fermions. The kinetic terms are canonical and the interactions
are given by\foot{The beta functions for all couplings in this theory can
be found at \url{http://het.ucsd.edu/misc/4D_betas2s12f.m}.}
\begin{multline*}
V=\tfrac{1}{24}\lambda_1\phi_1^4+\tfrac{1}{24}\lambda_2\phi_2^4
+\tfrac{1}{4}\lambda_3\phi_1^2\phi_2^2+\tfrac{1}{6}\lambda_4\phi_1^3\phi_2
+\tfrac{1}{6}\lambda_5\phi_1\phi_2^3\\
+\left(\phi_1\begin{pmatrix}\psi_1&\psi_2\end{pmatrix}\begin{pmatrix}y_1&y_2\\
y_3&y_4\end{pmatrix}\begin{pmatrix}\tilde{\psi}_1\\\tilde{\psi}_2\end{pmatrix}
+\phi_2\begin{pmatrix}\psi_1&\psi_2\end{pmatrix}\begin{pmatrix}y_5&y_6\\
y_7&y_8\end{pmatrix}\begin{pmatrix}\tilde{\psi}_1\\
\tilde{\psi}_2\end{pmatrix}+\text{h.c.}\right).
\end{multline*}
In contrast with the active Weyl spinors, the sterile ones do not interact
with the scalars, but they do interact with the gluons through their
kinetic terms. One needs sterile fermions in order to get a perturbative
fixed point for the gauge coupling, \textit{\`a la} Banks--Zaks
\cite{Banks:1981nn,Fortin:2011sz}. The smallest value of $\varepsilon$ for
which our theory is physical is $\varepsilon=\tfrac13$, but we will treat
$\varepsilon$ as an expansion parameter and take $\varepsilon\to\tfrac13$
at the end. As we will see, our perturbative results can be trusted in this
limit.

The most general virial current in our theory is\foot{Lower case indices
from the beginning of the roman alphabet are indices in flavor space for
scalar fields, while lower case indices from the middle are indices in
flavor and gauge space for Weyl spinors.}
\eqn{V^\mu=Q_{ab}\phi_a\partial^\mu\phi_b
-P_{ij}\bar{\psi}_ii\bar{\sigma}^\mu\psi_j\,,
}[CandidateV]
where $Q_{ab}$ is antisymmetric and $P_{ij}$ anti-Hermitian, i.e.,
$Q_{ba}=-Q_{ab}$ and $P^*_{ji}=-P_{ij}$. For compactness we have denoted by
$\psi_{3,4}$ the two antifundamentals $\tilde\psi_{1,2}$.
By gauge invariance $P_{ij}=P_{ji}=0$ for $i=1,2$ and $j=3,4$.  All these
constraints are satisfied by
\eqn{Q=\begin{pmatrix}
  0 & q\\
  -q & 0
\end{pmatrix}
\qquad\text{and}\qquad
P=
\begin{pmatrix}
  ip_1 & p_5 + ip_6 & 0 & 0 \\
  -p_5 + ip_6 & ip_2 & 0 & 0 \\
  0 & 0 & ip_3 & p_7  +ip_8 \\
  0 & 0 & -p_7 + ip_8 & ip_4
\end{pmatrix}.}[VirialCoeffs]
The virial current \CandidateV contains a fermionic part, something that
can lead to an ABJ-like anomaly \cite{Adler:1969gk,Bell:1969ts} for the
dilatation current. In this case, the virial current would have an extra
contribution to its anomalous dimension,\foot{As is also the case, e.g.,
for the axial current \cite[Appendix C]{Kaplan:1988ku}.} beyond the one
calculated from its two-point function. This is not allowed by conformal
invariance \cite{Fortin:2012hn}. Therefore, we expect that a limit cycle
solution should have the property that the virial current be not anomalous.
Consequently, a consistent limit cycle in a gauge theory with an
$SU(n\geq3)$ gauge group and fundamental and antifundamental fermions
should have
\eqn{\Tra P=0.}
This condition provides a powerful check on our computations.

\subsec{The three-loop computation}
It is convenient to rewrite compactly the interactions in $V$ as
\eqn{V=\tfrac1{4!}\lambda_{abcd}\phi_a\phi_b\phi_c\phi_d +
(\tfrac12 y_{a|ij}\phi_a\psi_i\psi_j+\text{h.c.}).}
Here, again, we are using the compact notation for the Weyl spinors, with
$\psi_{3,4}$ standing for $\tilde\psi_{1,2}$.  To find a limit cycle we
must exhibit solutions to
\eqna{\beta^g(g,y,\lambda)&=0\,,\\
\beta_{a|ij}(g,y,\lambda)&=-Q_{a'a}y_{a'|ij}-P_{i'i}y_{a|i'j}-
P_{j'j}y_{a|ij'}\,,\\
\beta_{abcd}(g,y,\lambda)&=-Q_{a'a}\lambda_{a'bcd}-Q_{b'b}\lambda_{ab'cd}
-Q_{c'c}\lambda_{abc'd}-Q_{d'd}\lambda_{abcd'}\,,}[SolnScale]
that do not require zero $\beta_{a|ij}$ and/or $\beta_{abcd}$. This
requires both determining the values of the coupling constants and of the
matrices $Q$ and $P$ for which the equations are satisfied.  It would
appear, naively, that the system of equations \SolnScale has more unknowns
than equations, due to the presence of the unknowns $Q_{ab}$ and $P_{ij}$,
and is thus ill-defined.  However, in searching for particular solutions,
one is free to set some coupling constants to zero. This is accomplished by
using the freedom to redefine the scalar fields by an $O(2)$ transformation
and the active Weyl spinors by a $U(2)\times U(2)$ transformation, with the
concomitant redefinition of coupling constants.  Note that a coupling may
become zero without its beta function becoming zero, since the couplings
are not exclusively multiplicatively renormalized. Hence, the number of
unknowns in \SolnScale is reduced and we obtain a well-defined system with
equal numbers of equations and unknowns.

As in Ref.~\cite{Fortin:2012ic} we can calculate the entries of $Q$ and $P$
on a limit cycle in an expansion in $\varepsilon$. To that end, we expand
in the small parameter $\varepsilon$ the couplings,
\eqn{g=\sum_{n\geq1}g^{(n)}\varepsilon^{n-\frac{1}{2}},\qquad
y_{a|ij}=\sum_{n\geq1}y_{a|ij}^{(n)}\varepsilon^{n-\frac{1}{2}},\qquad
\lambda_{abcd}=\sum_{n\geq1}\lambda_{abcd}^{(n)}\varepsilon^n\,,}
and the unknown parameters in the virial current,
\eqn{Q_{ab}=\sum_{n\geq3}Q_{ab}^{(n)}\varepsilon^n,\qquad
P_{ij}=\sum_{n\geq3}P_{ij}^{(n)}\varepsilon^n\,,}
and we solve Eqs.~\SolnScale order by order in $\varepsilon$.  The lowest
order entries in $Q$ and $P$ are of order $\varepsilon^3$, for at lower
orders in $\varepsilon$, corresponding to one- and two-loop orders in
perturbation theory, the beta functions produce a gradient flow
\cite{Jack:1990eb}.

To establish a limit cycle we have to compute the
$\varepsilon^3$-order terms in the $\varepsilon$ expansion of the
parameters of the virial current. For a complete calculation of these we
need the two-loop beta function for the quartic coupling, the three-loop
beta function for the Yukawa coupling, and the four-loop beta function for
the gauge coupling. To see why, let us explain how the $\varepsilon$
expansion works.

The $\varepsilon$ expansion of the beta functions can be written
schematically as
\vspace{5pt}
\eqna{\smash{\frac{\beta^g}{g^2}}&=\smash{\sum_{n\geq1}}f_g^{(n)}
\varepsilon^{n+\frac{1}{2}}=f^{(1)}_g(g^{(1)},y^{(1)})\varepsilon^{3/2}+
f^{(2)}_g(g^{(1)},y^{(1)},\lambda^{(1)};g^{(2)},y^{(2)})\varepsilon^{5/2}\\
&\hspace{5cm}+f^{(3)}_g(g^{(1)},y^{(1)},\lambda^{(1)},g^{(2)},y^{(2)},
\lambda^{(2)}; g^{(3)},y^{(3)})\varepsilon^{7/2}+\cdots,\\
\beta^y&=\smash{\sum_{n\geq1}f_y^{(n)}}\varepsilon^{n+\frac{1}{2}}=
f^{(1)}_y(g^{(1)},y^{(1)})\varepsilon^{3/2}+
f^{(2)}_y(g^{(1)},y^{(1)},\lambda^{(1)};g^{(2)},y^{(2)})\varepsilon^{5/2}\\
&\hspace{5cm}+f^{(3)}_y(g^{(1)},y^{(1)},\lambda^{(1)},g^{(2)},y^{(2)},
\lambda^{(2)};g^{(3)},y^{(3)})\varepsilon^{7/2}+\cdots,\\
\beta^\lambda&=\sum_{n\geq1}f_\lambda^{(n)}\varepsilon^{n+1}=
f^{(1)}_\lambda(g^{(1)},y^{(1)};\lambda^{(1)})\varepsilon^{2}
+f^{(2)}_\lambda(g^{(1)},y^{(1)},\lambda^{(1)},g^{(2)},y^{(2)};
\lambda^{(2)})\varepsilon^{3}+\cdots.}
Note that the gauge-coupling beta function is divided by $g^2$.  This way
systems of equations obtained at a specific $\varepsilon$ order contain the
same coefficients in the $\varepsilon$ expansion of the couplings and can
thus be solved simultaneously.  All couplings, $f$'s, and beta functions
carry flavor indices which we omit for brevity. It is important to realize
that both the one- and the two-loop order of $\beta^g$ contribute to
$f_g^{(1)}$, for we are fixing the gauge coupling to a point \textit{\`a
la} Banks--Zaks. The first step is to simultaneously solve $f_g^{(1)}=0$
and $f_y^{(1)}=0$, a system of nonlinear equations from which we get a set
of solutions $\{(g^{(1)},y^{(1)})\}$.  Each solution in this set is then
used to solve $f^{(1)}_\lambda=0$, another system of nonlinear equations,
which also gives a set of solutions $\{\lambda^{(1)}\}$. At this point we
can discard solutions with complex $\lambda^{(1)}$'s---those correspond to
nonunitary theories---and construct the set of solutions
$\mathcal{S}=\{(g^{(1)},y^{(1)},\lambda^{(1)})\}$. The determination of the
unknowns in $f^{(n\ge2)}_x$ requires solving simultaneous linear equations,
and so we have a unique solution for each element of $\mathcal{S}$.  At the
next step we use solutions in $\mathcal{S}$ to simultaneously solve
$f_g^{(2)}=0$ and $f_{y}^{(2)}=0$ for the unknowns $g^{(2)}$ and $y^{(2)}$,
which are thus uniquely determined.  These are used in $f_\lambda^{(2)}$
from which $\lambda^{(2)}$ is determined, and then we consider $f_g^{(3)}$
and $f_y^{(3)}$. These two functions receive contributions from the
$(n\leq4)$-loop orders of $\beta^g$ and the $(n\leq3)$-loop orders of
$\beta^y$. At this level we also have to take $Q$ and $P$ into account,
i.e., we have to see if there are solutions in the set $\mathcal{S}$ that
can lead to solutions of the linear equations $f_g^{(3)}=0$ and
$f_y^{(3)}=Qy+Py$ with $Q$ and/or $P$ nonzero. An indication of which
solutions in $\mathcal{S}$ may lead to non-vanishing $Q$ or $P$ is that,
already at the previous order, the beta functions for the coupling
constants that were set to zero do not vanish.

Now, the two-loop Yukawa and scalar coupling beta functions and the
three-loop gauge beta function can be found in the literature
\cite{Jack:1984vj, Pickering:2001aq}.   To establish the non-vanishing of
$Q$ or $P$ at lowest order, $\varepsilon^3$, the three-loop Yukawa beta
function and the four-loop gauge beta function are required. Fortunately, a
complete calculation of these beta functions is not needed. We parametrize
the beta functions at these orders by summing all possible monomials of
coupling constants of appropriate order with arbitrary coefficients $c_n$.
Then, by solving the set of linear equations $f_g^{(3)}=0$ and
$f_y^{(3)}=Qy+Py$, we determine which of these coefficients are involved in
the determination of $Q$ and $P$. There is a one-to-one correspondence
between each monomial in the beta functions and a three- or four-loop
Feynman diagram. Thus, rather than computing  some 1200 three-loop diagrams
for the Yukawa beta function and a larger number of four-loop diagrams for
the gauge beta function, we find that only a small number of diagrams needs
to be computed.

\begin{figure}[H]
  \centering
  \includegraphics[scale=0.91]{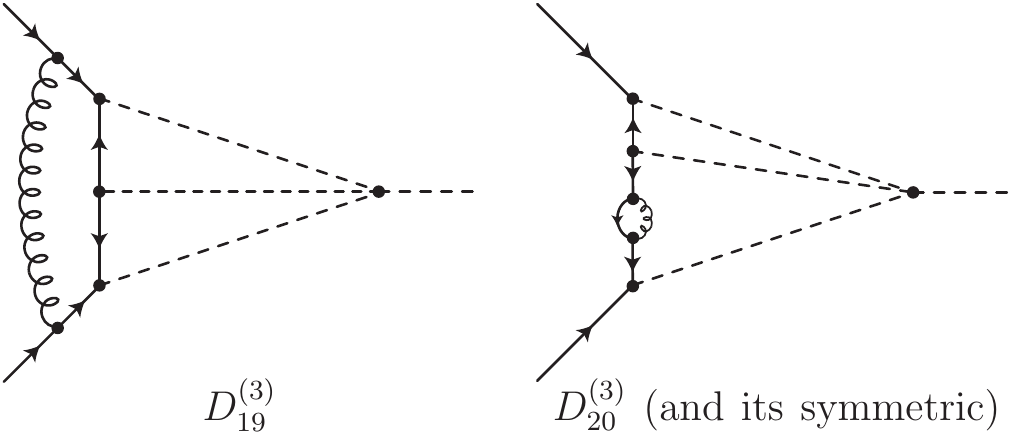}
  \caption{Three-loop diagrams that contribute to $q^{(3)}$ and
    $p^{(3)}$.}
  \label{ThreeLDiags}
\end{figure}

For the present model, following the procedure outlined in the previous
paragraph, we find, to lowest order in $\varepsilon$, that the point
\eqna{y_1&=\tfrac{\num{219}\sqrt{\num{92534}(\num{11430301}-\num{30212}
\sqrt{\num{19370}})}}{\num{27828258757}}
\pi\varepsilon^{3/2}+i\tfrac{24\sqrt{\num{3559}}}{\num{3559}}\pi
\sqrt{\varepsilon}+\cdots,\\
\qquad y_4&=\tfrac{8\sqrt{\num{17795}}}{\num{3559}}\pi\sqrt{\varepsilon}
+\cdots, \qquad
y_5=\tfrac{16\sqrt{\num{10677}}}{\num{3559}}\pi\sqrt{\varepsilon}+\cdots,\\
\lambda_1&=\tfrac{-3(\num{4177004}+\num{11781}\sqrt{\num{19370}})}
{\num{7819123}}\pi^2\varepsilon+\cdots,\qquad
\lambda_2=\tfrac{-75(\num{93964}+\num{1245}\sqrt{\num{19370}})}{\num{7819123}
}\pi^2\varepsilon+\cdots,\\
\lambda_3&=\tfrac{\num{1743}(9\sqrt{\num{19370}}-\num{676})}{\num{7819123}}
\pi^2\varepsilon+\cdots,\qquad
\lambda_4=\tfrac{-\num{249}\sqrt{78(\num{11430301}-\num{30212}\sqrt{\num{19370}})}}
{\num{7819123}}\pi^2\varepsilon+\cdots,\\
\lambda_5&=\tfrac{-63\sqrt{78(\num{11430301}-\num{30212}\sqrt{\num{19370}})}}
{\num{7819123}}
\pi^2\varepsilon+\cdots,\qquad
g=\tfrac{6\sqrt{\num{78298}}}{\num{3559}}\pi\sqrt{\varepsilon}+\cdots,}[SIpoint]
where we omit couplings that are zero at this point, lies on a limit cycle.
Among the zero couplings only the imaginary part of $y_5$ and the real part
of $y_8$ have nonzero beta functions and are thus generated on the limit
cycle.  Since not all imaginary parts of $y_{1,\ldots,8}$ can be rotated
away, the theory violates CP.  For the entries of $Q$ and $P$ we find
\eqna{q^{(3)}&=\frac{3\sqrt{\num{891563478}
-\num{2356536}\sqrt{\num{19370}}}}{\num{3763549370814194}} (\num{2061664} +
\num{143986}c_1 + \num{127268}c_2\\
&\quad - \num{735868}c_3 + \num{63634}c_4 - \num{735868}c_5 -
\num{1117968}c_6 - \num{1593120}c_7\\
&\quad + \num{654696}c_8 + \num{1309392}c_9\ + \num{1726320}c_{10}
+ \num{2146752}c_{11} - \num{25316928}c_{12}\\
&\quad + \num{24431904}c_{13} - \num{863136}c_{14} + \num{4779648}c_{15}
+ \num{106491}c_{16}\\
&\quad - \num{212982}c_{17} + \num{212982}c_{18}
+ \num{106491}c_{19} - \num{212982}c_{20}),}[qThreeLoop]
and
\eqna{p_1^{(3)}&=-\frac{18\sqrt{\num{297187826}
- \num{785512}\sqrt{\num{19370}}}}{\num{1881774685407097}}(\num{389632}
+ \num{4300}c_1 + \num{50720} c_2 - \num{105124}c_3\\
&\quad + \num{25360}c_4 - \num{105124}c_5 - \num{94632}c_6 - \num{357744}c_7
+ \num{93528}c_8 + \num{187056}c_9\\
&\quad + \num{276648}c_{10} + \num{276648}c_{11} - \num{3616704}c_{12}
+ \num{3490272}c_{13} - \num{155844}c_{14}\\
&\quad + \num{862992}c_{15} + \num{15213}c_{16}- \num{30426}c_{17}
+ \num{30426}c_{18} + \num{15213}c_{19} -\num{30426}c_{20})
- p_3^{(3)},}[pThreeLoop]
where the coefficients $c_{1,\ldots,20}$ are given by the contributions of
the three-loop diagrams of Fig.~\ref{ThreeLDiags} to the Yukawa beta
function.  None of the three- or four-loop contributions to $\beta^g$
appears in $q^{(3)}$ or $p^{(3)}_1$. For the other entries of $P$ we find
$p_{5,6,7,8}^{(3)}=0$, and that $p^{(3)}_{2,3,4}$ are undetermined with
$p^{(3)}_4=-p^{(3)}_2$.\foot{Undetermined entries of $P$ multiply operators
that are conserved, i.e., they correspond to global symmetries in the
fermionic sector of the theory.} The condition for absence of anomalies of
the dilatation current, $\Tra P=0$, is thus $p_1^{(3)}+p^{(3)}_3=0$. We
remark that $q^{(3)}$ and $p_1^{(3)}$ can be determined simply because the
running couplings $\Ima y_5$ and $\Rea y_8$ run through zero at the point
\SIpoint.
\begin{figure}[H]
  \centering
  \includegraphics[scale=0.91]{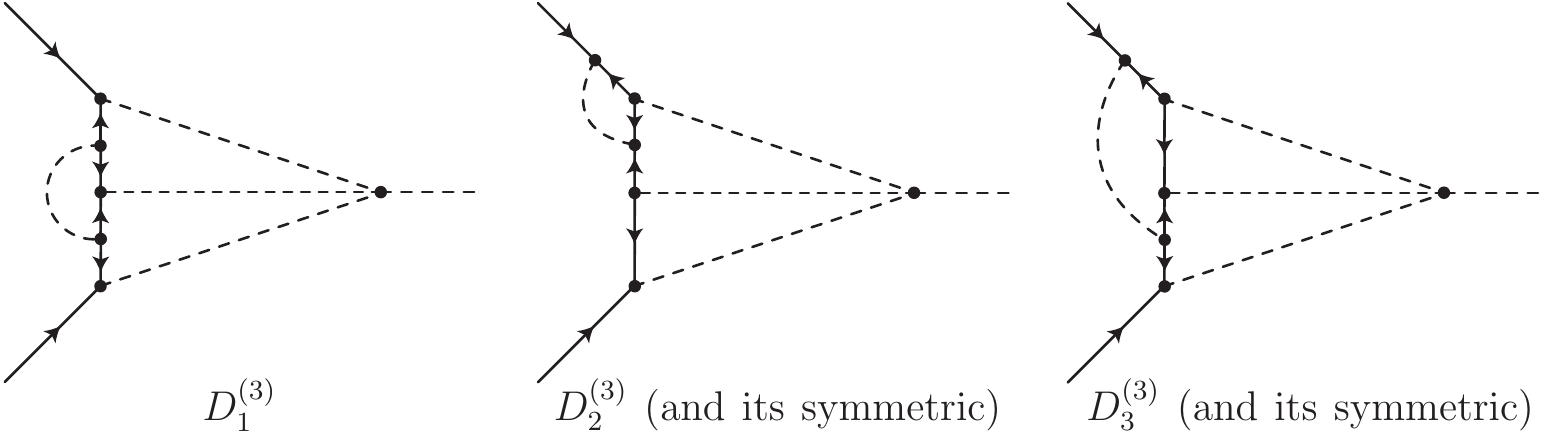}
\end{figure}
\begin{figure}[H]
  \centering
  \includegraphics[scale=0.91]{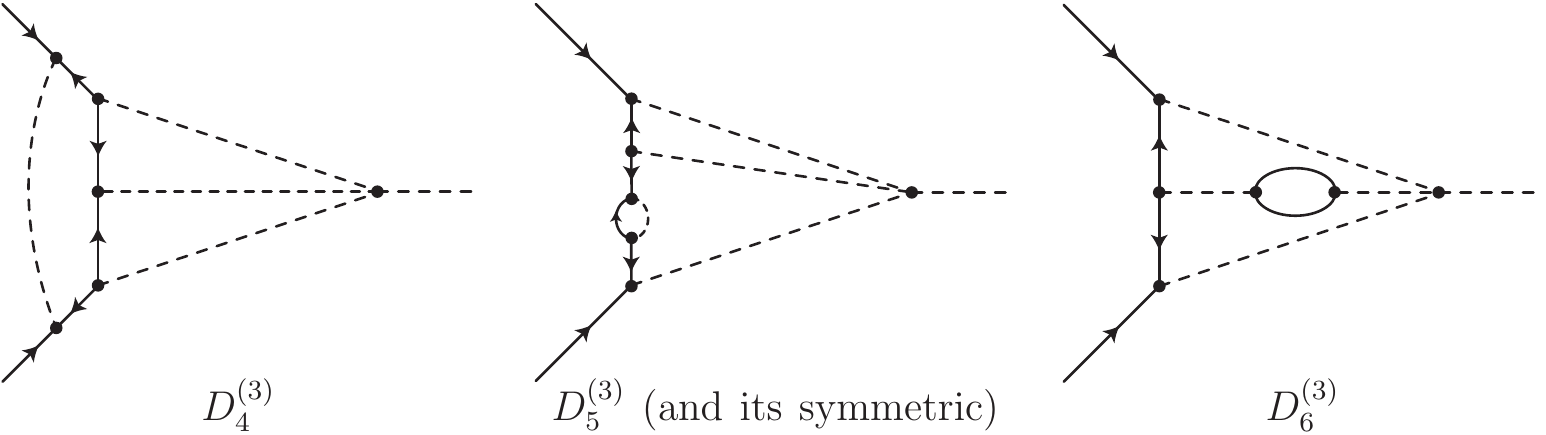}
\end{figure}
\begin{figure}[H]
  \centering
  \includegraphics[scale=0.91]{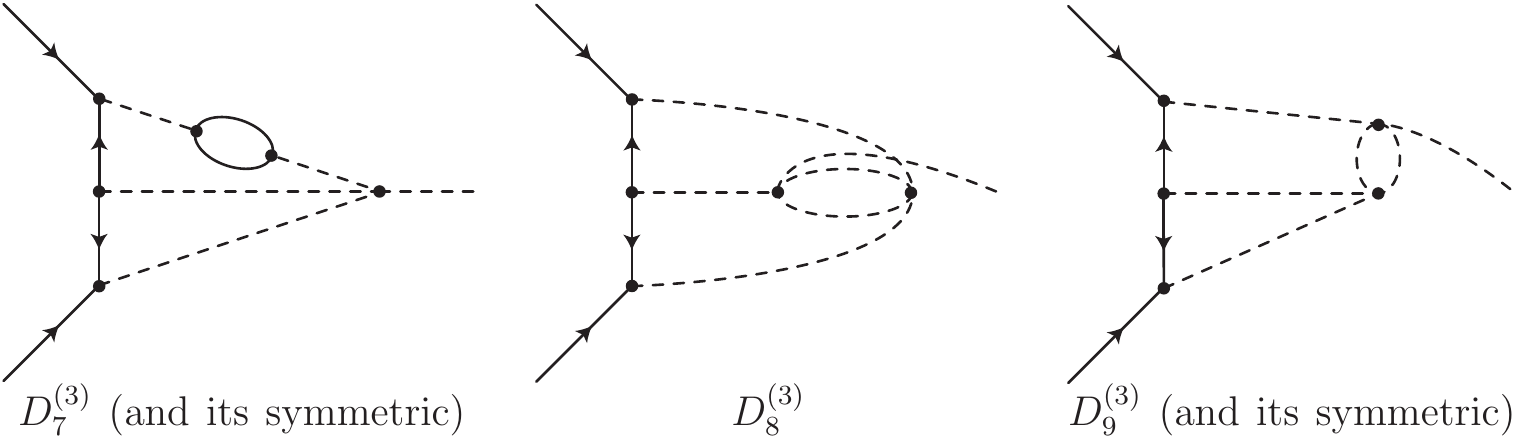}
\end{figure}
\begin{figure}[H]
  \centering
  \includegraphics[scale=0.91]{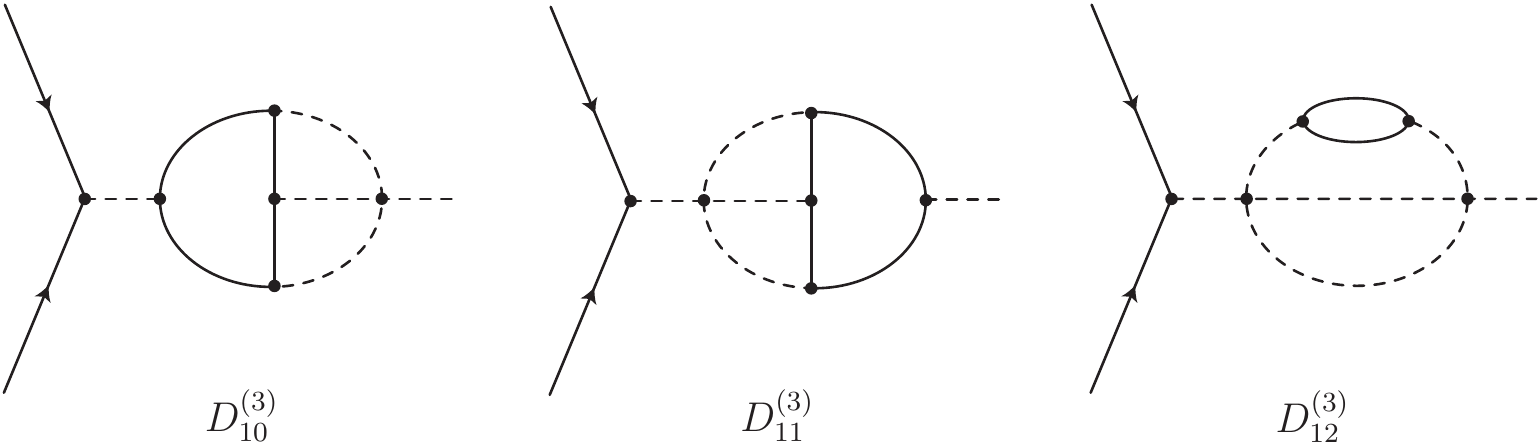}
\end{figure}
\begin{figure}[H]
  \centering
  \includegraphics[scale=0.91]{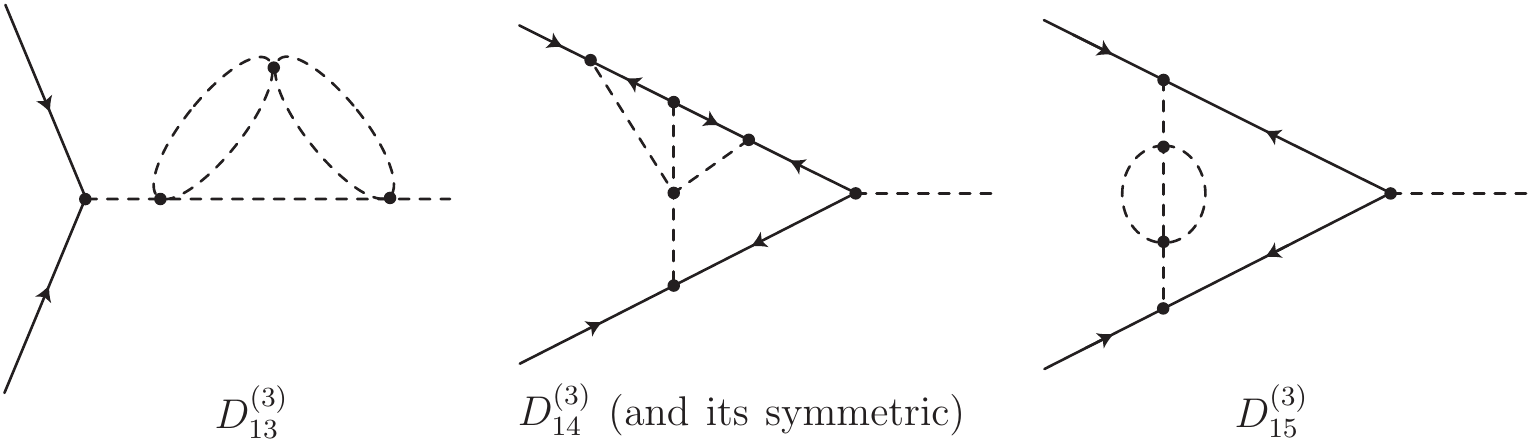}
\end{figure}
\begin{figure}[H]
  \centering
  \includegraphics[scale=0.91]{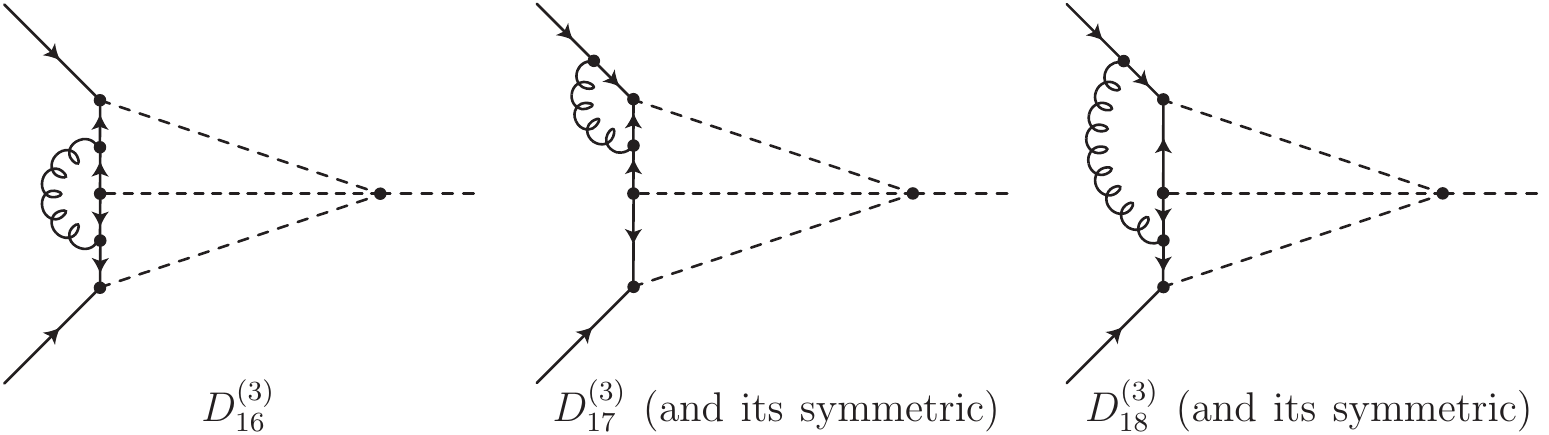}
\end{figure}

Note that both $q^{(3)}$ and $p_1^{(3)}$ receive contributions from exactly
the same diagrams, although with different weights, and that twelve of
these diagrams ($D^{(3)}_1$--$D^{(3)}_{10}$, $D^{(3)}_{12}$, and
$D^{(3)}_{13}$) are exactly the diagrams that contributed to the frequency
of the cycle of Ref.~\cite{Fortin:2012ic}. All diagrams involve at least
two types of couplings, as expected from the ``interference'' arguments of
Wallace and Zia \cite{Wallace:1974dy},  as  was also seen in our
three-loop calculations in $d=4-\epsilon$ spacetime dimensions
\cite{Fortin:2012ic}.

In dimensional regularization with $d=4-\epsilon$ the three-loop diagrams
of Fig.~\ref{ThreeLDiags} have simple $\epsilon$-poles and thus they
contribute to the Yukawa beta function. The residues of the simple
$\epsilon$-poles of $D_{1\text{--}20}^{(3)}$ lead to the coefficients
$c_{1,\ldots,20}$ in
\eqn{(16\pi^2)^3\beta_{a|ij}\supset c_1
(y^{\phantom{*}}_by^*_cy^{\phantom{*}}_d y^*_c
y^{\phantom{*}}_e)^{\phantom{*}}_{ij}\lambda_{abde}+\cdots
+c_{20}g^2[(y^{\phantom{*}}_by^*_c
t^{*A}t^{*A}y^{\phantom{*}}_d)^{\phantom{*}}_{ij}
+\{i\leftrightarrow j\}]\lambda_{abcd},}
as explained, for example, in \cite{Machacek:1983tz}. We performed the
three-loop computation with the method developed in
Ref.~\cite{Chetyrkin:1997fm} and the results of Ref.~\cite{Dreiner:2008tw}.
Since $q^{(3)}$ and $p_1^{(3)}$ are gauge-invariant, we can easily
incorporate a quick check in our calculation by using the full gluon
propagator, with the gauge parameter $\xi$. We find\foot{The symmetry
factors are included in the $c$'s.  Diagrams $D_{6\text{--}11}^{(3)}$ have
symmetry factor $s=\tfrac12$, diagrams $D_{12,13}^{(3)}$ have $s=\tfrac14$,
and diagram  $D_{15}^{(3)}$ has $s=\tfrac16$. All other diagrams have
$s=1$.}
\eqna{c_1&=3, & c_2&=-1, & c_3&=2, & c_4&=5, & c_5&=\tfrac12, &
c_6&=\tfrac32,\\
c_7&=\tfrac12, & c_8&=\tfrac32, & c_9&=\tfrac12, &
c_{10}&=\tfrac58, & c_{11}&=\tfrac58, & c_{12}&=-\tfrac{5}{32}\\
c_{13}&=-\tfrac{1}{16}, & c_{14}&=3, & c_{15}&=-\tfrac38, &
c_{16}&=-7+3\xi, & c_{17}&=4\xi, & c_{18}&=-7-\xi,\\
&&&& c_{19}&=19+5\xi, & c_{20}&=-\xi.}
Inserting these into the expressions \qThreeLoop and \pThreeLoop we obtain
\eqn{q^{(3)}=\frac{\num{20745} \sqrt{\num{891563478}-\num{2356536}
\sqrt{\num{19370}}}}{\num{99040772916163}}\approx 5\times 10^{-6},}
and
\eqn{p_1^{(3)} = -p_3^{(3)}.}
That $q^{(3)}\ne0$ indicates that we have a limit cycle in the RG running
of a four-dimensional unitary, renormalizable, well-defined gauge theory.
This is the first example ever exhibited of such behavior.  As expected,
there is no $\xi$-dependence in the final answer. As expected, the
dilatation current is automatically non-anomalous. These are nontrivial
checks and lend credibility to our calculation. We have found in the same
theory a distinct second limit cycle, in another position in coupling
space, with exactly the same properties as the one we presented above.

We have verified that our results can be trusted in the $\varepsilon\to
\tfrac13$ limit. More specifically, the expansion parameters are bounded on
the cycle:  $|\lambda|/16\pi^2 < 5\%$, $|y|^2/16\pi^2 < 1\%$, and
$g^2/16\pi^2 = 0.46\%$.  Hence, they remain perturbative along the whole
cycle.

The only unsatisfactory feature of our example is the fact that, as can be
seen from Eqs.~\SIpoint, the tree-level scalar potential is unbounded from
below. Still the model can be studied in perturbation theory, since the
vacuum state $\phi=0$ is perturbatively stable and its non-perturbative
lifetime $\tau$ is exponentially long, $\ln(\tau)\sim
1/\text{max}(\lambda_a)$.  This is similar in spirit to perturbative
studies of renormalization for $\phi^3$ models in six dimensions. However,
we expect that four-dimensional limit cycles with bounded scalar potential
also exist. Our expectation is based on our results in $d=4-\epsilon$,
where by progressing from the simplest examples, which displayed unbounded
tree-level potentials, to more involved examples, we found limit cycles
with bounded tree-level scalar potentials \cite{Fortin:2011ks,
Fortin:2012ic}.  In any case, the behavior of the effective potential in
any of these theories remains an open question.

\newsec{Conclusion}[conclusion]
The existence of limit cycles brings to light a new facet of unitary
four-dimensional QFT.  Many new questions arise:
\begin{itemize}
  \item What is the nature of RG flows away from limit cycles? Are there
  flows to or from fixed points from or to cycles or ergodic trajectories?
  \item Are there limit cycles in supersymmetric theories?
  \item Are there limit cycles in $d=3$ and $d>4$? Are there strongly-coupled
    limit cycles in $d=3$ that correspond to the $\epsilon\to1$ limit of
    the $d=4-\epsilon$ perturbative models?
  \item Are there limit cycles one can be establish in more indirect ways, i.e.,
    without the need of three-loop computations?
  \item Are there new possibilities for beyond the standard model physics
    associated with limit cycles \cite{Fortin:2011bm}?
  \item What is the holographic description of limit cycles? (This question
  has been considered in Refs.~\cite{Nakayama:2011zw, Nakayama:2012sn}.)
  \item Are there applications for condensed matter systems?
\end{itemize}
Answers to these questions will allow a more complete understanding of QFT,
and may lead to a new class of phenomena with unique characteristics. It
should already be clear, though, that RG flows display behavior that is
much richer than previously thought.

\ack{For our calculations we used the \emph{Mathematica} package
\href{http://www.feynarts.de/}{\texttt{FeynArts}}, and the programming
languages \href{http://www.python.org/}{\texttt{Python}} and
\href{http://www.nikhef.nl/~form/}{\texttt{FORM}}~\cite{Kuipers:2012rf}.
This work was supported in part by the US Department of Energy under
contract DOE-FG03-97ER40546.}

\bibliographystyle{JHEP}
\bibliography{LimitCyclesIn4d_ref}

\end{document}